\documentclass[%
 reprint,
 amsmath,amssymb,
 aps,
floatfix,
]{revtex4-2}

\usepackage{float}
\usepackage{setspace}
\usepackage{graphicx}% Include figure files
\usepackage{dcolumn}% Align table columns on decimal point
\usepackage{bm}% bold math
\usepackage{color} 
\begin{document}

\preprint{APS/123-QED}

\title{A comparative study between discrete and continuum models for the evolution of competing phenotype-structured cell populations in dynamical environments}

\author{Aleksandra Arda\v{s}eva}
 \email{aleksandra.ardaseva@maths.ox.ac.uk}
 \affiliation{Wolfson Centre for Mathematical Biology, University of Oxford, UK}%Lines break automatically or can be forced with \\
 
\author{Robert A. Gatenby}%
\affiliation{Department of Integrated Mathematical Oncology, H. Lee Moffitt Cancer Center, USA}

\author{Alexander R. A. Anderson}%
\affiliation{Department of Integrated Mathematical Oncology, H. Lee Moffitt Cancer Center, USA}

\author{Helen M. Byrne}%
\affiliation{Wolfson Centre for Mathematical Biology, University of Oxford, UK}

\author{Philip K. Maini}%
\affiliation{Wolfson Centre for Mathematical Biology, University of Oxford, UK}

\author{Tommaso Lorenzi}%
 \email{tommaso.lorenzi@polito.it}
\affiliation{Department of Mathematical Sciences ``G. L. Lagrange'', Dipartimento di Eccellenza 2018-2022, Politecnico di Torino, IT}

\date{\today}% It is always \today, today,
             %  but any date may be explicitly specified

\begin{abstract}
Deterministic continuum models formulated in terms of non-local partial differential equations for the evolutionary dynamics of populations structured by phenotypic traits have been used recently to address open questions concerning the adaptation of asexual species to periodically fluctuating environmental conditions. These deterministic continuum models are usually defined on the basis of population-scale phenomenological assumptions and cannot capture adaptive phenomena that are driven by stochastic variability in the evolutionary paths of single individuals. In light of these considerations, in this paper we develop a stochastic individual-based model for the coevolution between two competing phenotype-structured cell populations that are exposed to time-varying nutrient levels and undergo spontaneous, heritable phenotypic variations with different probabilities. Here, the evolution of every cell is described by a set of rules that result in a discrete-time branching random walk on the space of phenotypic states, and  nutrient levels are governed by a difference equation in which a sink term models nutrient consumption by the cells. We formally show that the deterministic continuum counterpart of this model comprises a system of non-local partial differential equations for the cell population density functions coupled with an ordinary differential equation for the nutrient concentration. We compare the individual-based model and its continuum analogue, focussing on scenarios whereby the predictions of the two models differ. The results obtained clarify the conditions under which significant differences between the two models can emerge due to stochastic effects associated with small population levels. In particular, these differences arise in the presence of low probabilities of phenotypic variation, and become more apparent when the two populations are characterised by less fit initial mean phenotypes and smaller initial levels of phenotypic heterogeneity. The agreement between the two modelling approaches is also dependent on the initial proportions of the two populations.
\end{abstract}

%\keywords{Suggested keywords}%Use showkeys class option if keyword
                              %display desired
\maketitle

\section{\label{intro}Introduction}
Adaptation to dynamically changing environments occurs in a variety of biological and ecological contexts~\citep{gillies2018eco, kussell2005bacterial, kremer2013coexistence, levins1968evolution}. In particular, when changes in nutrient availability occur, individuals in a population can either adopt a highly plastic phenotype~\citep{xue2018benefits}, which enables them to acquire different traits based on environmental cues, or a risk spreading strategy (\emph{e.g.} bet-hedging), which allows at least some fraction of the population to survive in the face of sudden environmental changes by producing offspring adapted to the new conditions~\citep{hopper1999risk, kussell2005phenotypic, philippi1989hedging}.

Mathematical modelling of evolutionary dynamics in time-varying environments has received considerable attention from mathematicians and physicists over the past fifty years -- see, for instance, \citep{baron2018successful,canino2019fluctuating,cvijovic2015fate,fuentes2017environmental,gomez2016evolutionary,gomez2019epigenetic,hiltunen2015environmental,jansen2005making,levin1976population,soyer2010evolution,tuljapurkar1982population,wienand2018eco} and references therein. Recently, deterministic continuum models formulated in terms of non-local partial differential equations (PDEs) for the evolutionary dynamics of populations, structured by phenotypic traits, have been used to address open questions concerning the adaptation of asexual species to periodically fluctuating environments~\cite{almeida2019evolution, ardavseva2019evolutionary, ardavseva2019mathematical, carrere2017influence, figueroa2018long, iglesias2019selection, lorenzi2015dissecting, mirrahimi2015time}. 

Although more amenable to analytical and numerical approaches, which allow for an in-depth theoretical understanding of the underlying dynamics, these deterministic continuum models are usually defined on the basis of population-scale phenomenological assumptions. This makes it more difficult to incorporate the finer details of phenotypic adaptation by single individuals. Moreover, such models cannot capture adaptive phenomena that are driven by stochastic effects in the evolutionary paths of single individuals. This will be particularly relevant at low population levels, which are commonly observed when risk-spreading adaptive strategies occur~\cite{muller2013bet}. Ideally, we want to derive deterministic continuum models from first principles (\emph{i.e.} as the appropriate limit of discrete stochastic models that track the evolution of single individuals), which permit the representation of individual-scale adaptive mechanisms, and account for possible stochastic inter-individual variability in evolutionary trajectories~\citep{champagnat2006unifying, champagnat2002canonical, chisholm2016evolutionary, stace2019discrete}.

In light of these considerations, we develop a stochastic individual-based (IB) model for the evolutionary dynamics of two competing phenotype-structured cell populations that are exposed to time-varying nutrient levels and undergo spontaneous, heritable phenotypic variations with different probabilities. In this model, every cell is viewed as an individual agent whose phenotypic state is modelled by a discrete variable, which represents the normalised level of expression of a gene that allows cells to cope with nutrient scarcity. For instance, activation of hypoxia-inducible factors allows mammalian cells to adapt to oxygen deprivation~\cite{chan2007hypoxia}. In the model, cells proliferate, die and undergo phenotypic variations according to a set of rules that correspond to a discrete-time branching random walk on the space of phenotypic states~\citep{chisholm2016evolutionary,hughes1995random}. We assume that the cell proliferation rate depends on nutrient levels, and that nutrient concentration is governed by a difference equation in which a sink term models nutrient consumption by the cells.

We show formally that the deterministic continuum counterpart of this stochastic IB model comprises a system of non-local PDEs for the cell population density functions ({\it i.e.} the cell distribution over the space of phenotypic states) coupled with an ordinary differential equation for the nutrient concentration. Such a continuum model is analogous to the models that we have previously studied analytically and numerically in~\cite{ardavseva2019evolutionary,ardavseva2019mathematical}. Moreover, we carry out a comparative study between the IB model and its continuum analogue, to explore scenarios in which differences between the two models emerge due to stochastic effects not captured by the deterministic continuum model. 
%This would make it possible to explore the impact of stochastic fluctuations in single-cell phenotypic properties on the outcome of the competition between cell populations undergoing phenotypic variations at different rates. Such stochastic effects are expected to be relevant in the regime of low cell numbers and cannot easily be captured by continuum models like the one considered here.

The paper is organised as follows. In Section~\ref{model}, we introduce the stochastic IB model. In Section~\ref{modelc}, we present its deterministic continuum counterpart (a formal derivation is provided in Appendix~\ref{formalderiv}). In Section~\ref{results}, we present the main results of the comparative study of the two models. These results are discussed in Section~\ref{conclusion}, which concludes the paper and provides a brief overview of possible research perspectives.

\section{\label{model}Stochastic individual-based model}
We model the evolutionary dynamics of two competing cell populations in a well-mixed system. Cells in the two populations proliferate (\emph{i.e.} divide), die and undergo spontaneous, heritable phenotypic variations. We assume that the two populations differ only in their probability of phenotypic variation. The population undergoing phenotypic variations with a higher probability is labelled by the letter $H$, while the other population is labelled by the letter $L$. The phenotypic state of every cell at time $t \in [0,t_f] \subset \mathbb{R}^+$ is characterised by a variable $x \in [0,1] \subset \mathbb{R}^+$, which represents the normalised level of expression of a gene that allows cells to cope with nutrient deprivation. In particular, we assume that cells in the phenotypic state $x=0$ are best adapted to nutrient-rich environments, whereas cells in the phenotypic state $x=1$ are best adapted to nutrient-scarce environments.

We represent each cell as an agent that occupies a position on a lattice. We discretise the time variable and the phenotypic state via $t_{h} = h \tau \in [0,t_f]$ and  $x_{j} = j  \chi \in [0,1]$, respectively, where $h, j \in \mathbb{N}_{0}$, and $\tau \in \mathbb{R}^+_*$ and $\chi \in \mathbb{R}^+_*$ are the time- and phenotype-step, respectively. We introduce the dependent variable $N^{h}_{i,j}\in\mathbb{N}_0$ to represent the number of cells of population $i \in \{H, L\}$ on lattice site $j$ (\emph{i.e.} in the $j^{th}$ phenotypic state) at time-step $h$. The  density (\emph{i.e.} the phenotype distribution) of population $i$, the size of population $i$, and the total number of cells are defined, respectively, as follows
\begin{equation}
\label{e:SP1a}
n_i(t_h,x_{j})= n^h_{i,j} := N^{h}_{i,j} \,  \chi^{-1},
\end{equation}
\begin{equation}
\label{e:SP1b}
\rho_i(t_h)=\rho^h_i := \sum_j N^{h}_{i,j} \; \text{ and } \; \rho(t_h)=\rho^h := \sum_i \rho^h_i.
\end{equation}
We further define the mean phenotype of population $i$ and the related standard deviation, respectively, as
\begin{equation}
\label{e:SP2}
\mu_i(t_h) = \mu^h_i := \frac{1}{\rho^h_i} \sum_{j} x_j \, N^{h}_{i,j}
\end{equation}
and
\begin{equation}
\label{e:SP3}
\sigma_i(t_h) = \sigma^{h}_i := \left(\frac{1}{\rho^h_i} \sum_{j} x^2_j \, N^{h}_{i,j}  -  \left(\mu^h_i\right)^2 \right)^{\frac{1}{2}}.
\end{equation}
Finally, the nutrient concentration at time-step $h$ is modelled by the discrete, non-negative function $S(t_h) = S^{h}$.

\subsection{Mathematical modelling of phenotypic variation}
We account for spontaneous, heritable phenotypic variation by allowing cells to update their phenotypic states according to a random walk. More precisely, between the time-steps $h$ and $h+1$, every cell in population $i \in \{H,L \}$ either enters a new phenotypic state, with probability $\lambda_i \in [0,1]$, or remains in its current phenotypic state, with probability $1-\lambda_i$. Since we assume phenotypic variations occur randomly due to non-genetic instability, rather than selective pressures~\cite{huang2013genetic}, then a cell of population $i$ in phenotypic state $x_j$ that undergoes phenotypic variation enters into either of the phenotypic states $x_{j\pm1} = x_{j} \pm \chi$ with probabilities $\lambda_i/2$. No-flux boundary conditions are implemented by aborting any attempted phenotypic variation of a cell if it requires moving into a phenotypic state outside the interval $[0,1]$.

\subsection{Mathematical modelling of cell division and death}
Cells divide, die or remain quiescent with probabilities that depend on their phenotypic states, the total number of cells and the nutrient concentration. We assume that a dividing cell is replaced by two identical cells that inherit the phenotypic state of the parent cell (\emph{i.e.} the progenies are placed on the same lattice site as their parent), while a dying cell is removed from the population. 

In order to translate into mathematical terms the idea that larger population sizes correspond to more intense competition between cells, at every time-step $h$ we allow cells to die due to intra-population and inter-population competition at a rate proportional to the total cell number $\rho^h$, with constant of proportionality $d > 0$.

We denote by $p(x_j,S^h)$ the division rate of a cell in the $j^{th}$ phenotypic state, where $S^h$ is the nutrient concentration. Since $x_j$ represents the normalised expression level of a gene that allows cells to cope with nutrient scarcity, we assume that phenotypic variants with $x_j \to 0$ are characterised by the maximal division rate when nutrient is abundant (\emph{i.e.} if $S^h \to \infty$), whereas phenotypic variants with $x_j \to 1$ are characterised by the maximal division rate when nutrient is scarce (\emph{i.e.} if $S^h \to 0$). Our implicit assumption here is that cells in the phenotypic state $x=1$ switch to other nutrients that are abundant, and therefore they are no longer dependent on the specific nutrient we are modelling. For example, cancer cells are known to consume glucose, an alternative but inefficient energy source, rather than oxygen~\cite{vander2009understanding}. In this example: cells in the phenotypic state $x=0$ would have a fully oxidative metabolism and would produce energy through oxygen consumption only; cells in the phenotypic state $x=1$ would express a fully glycolytic metabolism and would produce energy through glucose consumption only; cells in other phenotypic states $x \in (0,1)$ would produce energy via both oxygen and glucose consumption, and higher values of $x$ would correlate with a less oxidative and more glycolytic metabolism. Under these assumptions, and following the modelling strategies that we proposed in~\citep{ardavseva2019evolutionary, ardavseva2019mathematical}, we define the cell division rate $p(x_j,S^h)$ as follows:
\begin{multline}
\label{defp}
p(x_j,S^h) := \gamma \frac{S^h}{1+S^h}(1-x_j^2) \\+ \zeta \left(1 - \frac{S^h}{1+S^h}\right)\left[1-(1-x_j)^2\right].
\end{multline}
In~\eqref{defp}, the parameters $\gamma > 0$ and $\zeta > 0$ model, respectively, the maximum cell division rate of the phenotypic variants best adapted to nutrient-rich and nutrient-scarce environments (\emph{i.e.} cells in the phenotypic states $x_j=0$ and $x_j=1$, respectively). To incorporate into the model the possible fitness cost associated with the ability to survive in nutrient-scarce environments~\cite{hereford2009quantitative,basanta2008evolutionary}, we make the additional assumption that $\zeta \leq \gamma$.

After a little algebra, definition~\eqref{defp} can be rewritten as 
\begin{equation} \label{prewritten}
p(x_j,S^h) = \gamma \, g(S^h) - h(S^h) (x_j-\varphi(S^h))^2,
\end{equation}
where
$$
g(S^h) := \frac{1}{1+S^h} \left[S^h + \frac{1}{\dfrac{\gamma}{\zeta} \left(1 + \dfrac{\gamma}{\zeta} S^h\right)}\right], 
$$
\begin{equation}
 \label{def:phi} 
\varphi(S^h) := \frac{1}{1 + \dfrac{\gamma}{\zeta} S^h}
\end{equation}
and
$$
h(S^h) := \zeta \left[1 + \left(\frac{\gamma}{\zeta} - 1\right)\frac{S^h}{1+S^h}\right].
$$
Here, $\gamma g(S^h)$ is the maximum fitness, $\varphi(S^h)$ is the fittest phenotypic state and $h(S^h)$ is a selection gradient. Notice that, consistent with our modelling assumptions, $\varphi : [0,\infty) \to [0,1]$, $\displaystyle{\lim_{S\to0} \varphi(S) = 1}$ and $\displaystyle{\lim_{S\to\infty} \varphi(S) = 0}$. 

Under these assumptions, between time-steps $h$ and $h+1$ a cell in the $j^{th}$ phenotypic state may divide with probability
    \begin{equation} \label{pb}
        \mathcal{P}_b := \tau \, p(x_j,S^h),
    \end{equation}
die with probability    
    \begin{equation} \label{pd}
         \mathcal{P}_d := \tau \,d \, \rho^h,
    \end{equation}
or remain quiescent (\emph{i.e.} do not divide nor die) with probability
    \begin{equation} \label{pq}
         \mathcal{P}_q := 1- \tau \, \left(p(x_j,S^h)  + d \, \rho^h\right).
    \end{equation}
Notice that we are implicitly assuming that the time-step $\tau$ is sufficiently small that $0 < \mathcal{P}_i < 1$ for all $i \in \{b,d,q \}$.

\subsection{\label{nutrient} Mathematical modelling of nutrient dynamics}
%We consider a scenario where the evolution of $S^h$ is governed by a difference equation that is coupled to the cell dynamics so that we may investigate how the evolutionary dynamics of the cells are shaped by the negative feedback that regulates the growth of the two populations through nutrient consumption. 
Following~\citet{ardavseva2019mathematical}, we describe the nutrient dynamics via the following difference equation for $S^h$
\begin{multline}\label{Sode}
    S^{h+1} = S^{h} + \tau\Big[I^h - \eta S^h \\ - \theta \gamma \frac{S^h}{1+S^h} \sum_j (1-x_j)^2 \left(N^h_{H,j} + N^h_{L,j}\right) \Big],
\end{multline}
complemented with a suitable initial nutrient concentration $S^0$. Since we consider a well-mixed system, there is no diffusion of the nutrient. In~\eqref{Sode}, the parameter $\eta>0$ represents the rate of natural decay of the nutrient, while the last term on the right-hand side of~\eqref{Sode} models the rate of nutrient consumption by the cells and is based on the following argument. Cells in the phenotypic state $x_j=1$ do not rely on the nutrient we are modelling for their survival -- these cells might produce energy via different metabolic pathways that do not require the nutrient under consideration -- and, as such, they do not consume any nutrient. By contrast, cells in the phenotypic state $x_j=0$ consume the nutrient at a rate proportional to their cell division rate, with constant of proportionality $\theta>0$. Finally, the rate at which the nutrient is consumed by cells in phenotypic states $x_j \in (0,1)$ is a fraction of the consumption rate of cells in the phenotypic state $x_j=0$, with higher values of $x_j$ correlating with lower rates of nutrient consumption. The discrete, non-negative function $I^h$ on the right-hand side of~\eqref{Sode} models the rate at which the nutrient is supplied to the system. When the nutrient inflow is constant we fix
\begin{equation}
\label{Iconst}
I^h \equiv \bar{I} \geq 0;
\end{equation}
when the nutrient inflow undergoes periodic oscillations we prescribe
\begin{equation}
\label{Ifluc}
I^h := \textup{max}\left(0, A\sin \left(\frac{2\pi t_h}{T}\right)\right),
\end{equation}
with the parameters $T>0$ and $A>0$ modelling, respectively, the period and the amplitude of the oscillations. 

\subsection{Computational implementation}
\begin{figure}[h!]
\includegraphics[width=1\linewidth]{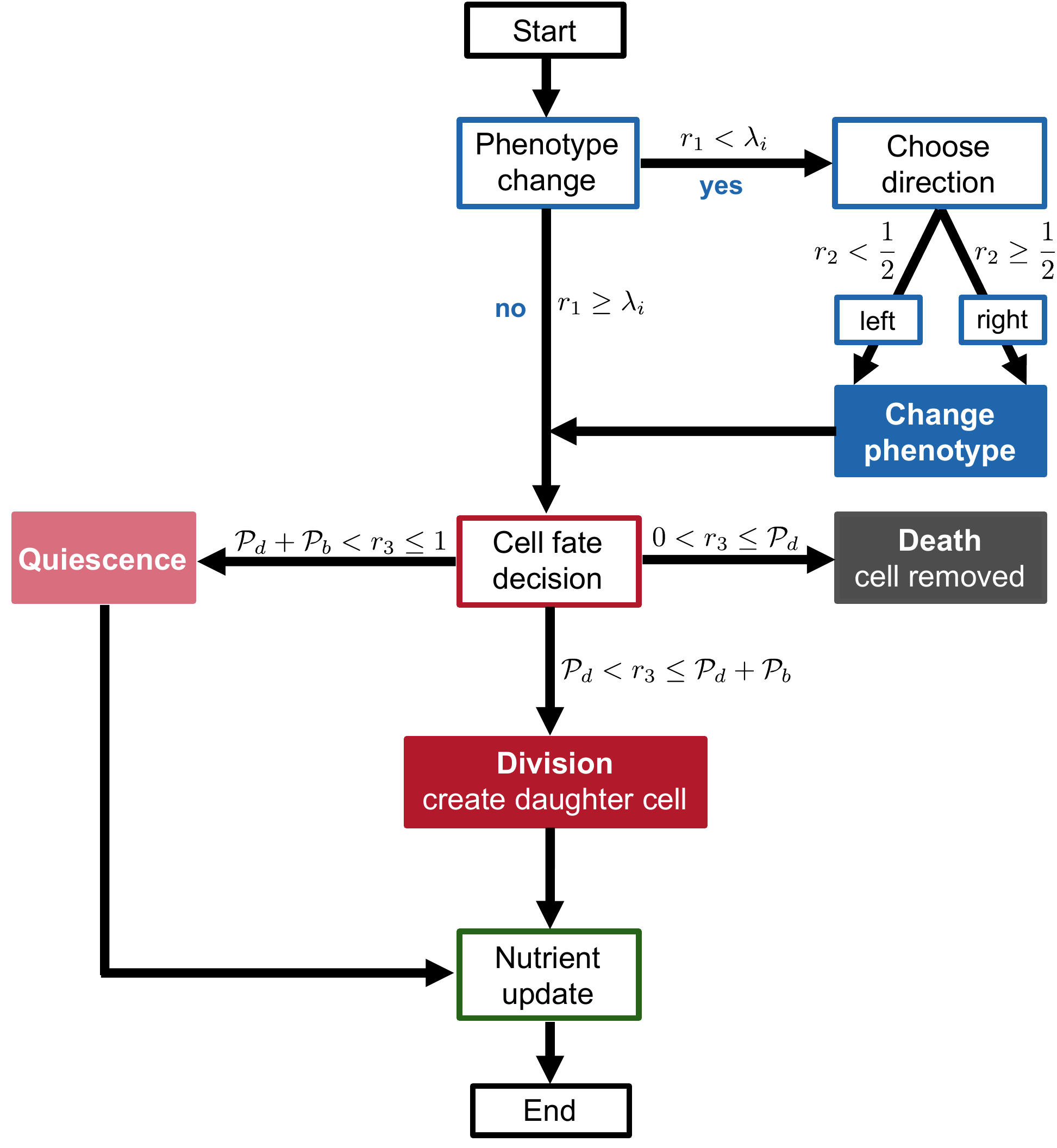}% Here is how to import EPS art
\caption{\label{fig1} {\bf Flowchart illustrating the procedure underlying the computational implementation of the stochastic IB model for each cell at every time-step}. Once all cells have undergone both the phenotype-change step and the fate-decision step, the total number of cells is computed and the nutrient level is updated.}
\end{figure}

Numerical simulations of the IB model are performed using the open-source Java library \texttt{Hybrid Automata Library (HAL)}~\citep{bravo2020hybrid}. At each time-step, we follow the procedures summarised in Figure~\ref{fig1} and described hereafter to simulate phenotypic variation as well as cell division and death. All random numbers mentioned below are real numbers drawn from the standard uniform distribution on the interval $(0,1)$ using the Java function \texttt{Rand.Double()}.
%We define the time step, $\tau$, and phenotype step, $\chi$, such that $\beta_i = \lambda_i/2$,
\\\\
\paragraph{Computational implementation of spontaneous, heritable phenotypic variation.} For each cell in population $i$, a random number, $r_1$, is generated and used to determine whether the cell undergoes a phenotypic variation (\textit{i.e.} $0 < r_1 < \lambda_i$) or not (\textit{i.e.}  $\lambda_i \leq r_1 <  1$). If the cell undergoes a phenotypic variation, then a second random number, $r_2$, is generated. If $0 < r_2 < 1/2$, then the cell moves into the phenotypic state to the left of its current state, {\it i.e.} a cell in the phenotypic state $x_j$ will move into the phenotypic state $x_{j-1}=x_{j}-\chi$, whereas if $1/2 \leq r_2 < 1$ then the cell moves into the phenotypic state to the right of its current state, {\it i.e.} a cell in the phenotypic state $x_j$ will move into the phenotypic state $x_{j+1}=x_{j}+\chi$. No-flux boundary conditions are implemented by aborting attempted phenotypic variations that would move a cell into a phenotypic state outside the unit interval.
\\\\
\paragraph{Computational implementation of cell division and death.} For each population, the number of cells in each phenotypic state is counted. The size of each cell population and the total number of cells are then computed via~\eqref{e:SP1b}. Equations~\eqref{pb}--\eqref{pq} are used to calculate the probabilities of cell division, death and quiescence for every phenotypic state. For each cell, a random number, $r_3$, is generated and the cells' fate is determined by comparing this number with the probabilities of division, death and quiescence corresponding to the phenotypic state of the cell. If $0 < r_3 < \mathcal{P}_d$ then the cell is considered dead and is removed from the population. If  $\mathcal{P}_d \leq r_3 < \mathcal{P}_d + \mathcal{P}_b$ then the cell undergoes division and an identical daughter cell is created. Finally, if $\mathcal{P}_d  + \mathcal{P}_b \leq r_3 < 1$ then the cell remains quiescent (\textit{i.e.} does not divide nor die).
\\\\
\paragraph{Computational implementation of nutrient dynamic.} At each time-step, the number of cells of the two populations in each phenotypic state is counted in order to evaluate the last term on the right-hand side of~\eqref{Sode}. The nutrient concentration is then updated via the difference equation~\eqref{Sode}.

\section{\label{modelc} Corresponding deterministic continuum model}
Using the formal method presented in~\cite{chisholm2016evolutionary, stace2019discrete}, we let the time-step $\tau \to 0$ and the phenotype-step $\chi \to 0$ in such a way that
\begin{equation}
\label{asymatdr}
\frac{\lambda_i \chi^2}{2 \tau} \rightarrow \beta_i \in \mathbb{R}^+_* \quad \text{ for } \quad i \in \{H, L\}.
\end{equation} 
Here, the parameter $\beta_i$ is the rate of spontaneous, heritable phenotypic variations of cells in population $i$. It is then possible to formally show (see Appendix~\ref{formalderiv}) that the deterministic continuum counterpart of the stochastic IB model is given by the following system of non-local PDEs for the cell population density functions $n_H(x,t)$ and $n_L(x,t)$ 
\begin{equation}
\label{eS1}
\left\{
\begin{array}{ll}
\displaystyle{\frac{\partial n_H}{\partial t} =  \beta_H \frac{\partial^2 n_H}{\partial x^2} + \big(p(x, S(t)) - d \, \rho(t)\big) n_H}, 
\\\\
\displaystyle{\frac{\partial n_L}{\partial t} = \beta_L \frac{\partial^2 n_L}{\partial x^2} + \big(p(x, S(t)) - d \, \rho(t)\big) n_L},
\\\\
\displaystyle{\rho(t) := \rho_H(t) + \rho_L(t), \;\; \rho_i(t) := \int_{0}^{1} n_{i}(x,t) \; {\rm d}x,}
\end{array}
\right.
\end{equation} 
posed on $(0,1) \times (0,t_f]$ and subject to zero-flux boundary conditions, \emph{i.e.}
\begin{equation}
\label{BCs}
\frac{\partial n_i(0,t)}{\partial x} = 0, \quad \frac{\partial n_i(1,t)}{\partial x} = 0 \quad \text{for all } t \in (0,t_f].
\end{equation} 
In~\eqref{eS1}, the nutrient concentration $S(t)$ is governed by the continuum counterpart of the difference equation~\eqref{Sode}, \emph{i.e.} the following integro-differential equation posed on $(0,t_f]$
%\begin{multline} 
%    \frac{\mathrm{d}S }{\mathrm{d} t} = I(t) - \lambda S - \\  \theta \gamma \frac{S}{1+S} \int_0^1 (n_H + n_L) (1-x^2) \; {\rm d}x,
%\end{multline}
\begin{equation}
\label{SodeCont}
\frac{\mathrm{d}S}{\mathrm{d} t} = I(t) - \lambda S - \\  \theta \gamma \frac{S}{1+S} \int_0^1 (1-x^2) \ (n_H + n_L) \; {\rm d}x,
\end{equation} 
which can be easily obtained in a formal way by letting $\tau \to 0$ and $h \to 0$ in~\eqref{Sode}. In the continuum modelling framework given by~\eqref{eS1}, the mean phenotype of population $i$ and the related standard deviation are defined, respectively, as
\begin{equation}
\label{e:SP2cont}
\mu_i(t) := \frac{1}{\rho_i(t)} \int_{0}^{1} x \, n_{i}(x,t) \; {\rm d}x
\end{equation} 
and
\begin{equation}
\label{e:SP3}
\sigma_{i}(t) := \left(\frac{1}{\rho_i(t)} \int_{0}^{1} x^2 \; n_i(x,t) \; {\rm d}x -  \mu_i^2(t) \right)^{\frac{1}{2}}.
\end{equation}

\section{\label{results}Main results}
In this section, we compare the results of numerical simulations of the stochastic IB model introduced in Section~\ref{model} and numerical solutions of the corresponding deterministic continuum model presented in Section~\ref{modelc}. 
\begin{table*}[!htb]
\caption{\label{table1} Parameter values used in numerical simulations.}
\begin{ruledtabular}
\begin{tabular}{ccc}
&Description& Values\\ \hline
$\lambda_H$ & Probability of phenotypic variation of population $H$ & [0.05, 1]\\
$\lambda_L$ & Probability of phenotypic variation of population $L$ & [0.02, 0.2] \\
$\gamma $& Maximum cell division rate of phenotypic variants best adapted to nutrient-abundant environments& 100\\
$\zeta$ & Maximum cell division rate of phenotypic variants best adapted to nutrient-scarce environments& 50 \\
$d$ & Death rate due to inter- and intra-population competition & 0.01\\
$\theta$ & Consumption rate of nutrient & [10$^{-5}$, 10$^{-3}$]\\
$\eta$ & Rate of natural decay of nutrient & $10^{-4}$\\
%$a_i$ & Initial size of population $i$& 800 \\
%$c_i$ & Initial mean phenotype of population $i$ & [0,1] \\
%$b_i$ & Initial inverse variance of population $i$ & [10,1000] \\
$\chi$ & Phenotype-step & 0.032 \\
$\tau$ & Time-step & $1.024 \times 10^{-3}$ \\
$S^0$ & Initial nutrient concentration & $10$\\
$t_f$ & Final time & [10, 40]
\end{tabular}
\end{ruledtabular}
\end{table*}

For consistency with previous mathematical studies of the evolutionary dynamics of phenotype structured populations, which rely on the prima facie assumption that population densities are Gaussians~\citep{rice2004evolutionary}, simulations are carried out under the assumption that the initial phenotype distribution of population $i$ for the IB model is of the form
\begin{equation} \label{initcondsdi}
n^0_{i,j} = a_i \left(\frac{b}{2\pi} \right)^{\frac{1}{2}} \exp \left[-\frac{b}{2} (x_j-c)^2 \right],
\end{equation}
with $i \in \{H, L\}$. In~\eqref{initcondsdi}, the parameter $a_i$ is related to the initial size of population $i$, while the parameters $b$ and $c$ are related, respectively, to the inverse of the initial standard deviation and the initial mean phenotype of the two populations. The initial population density $n_i(x,0)$ for the continuum model is defined as the continuum analogue of~\eqref{initcondsdi} (see Appendix~\ref{nummet}).

First, we present a sample of base-case results that demonstrate excellent quantitative agreement between the stochastic IB model and its deterministic continuum counterpart. Then, we perform a systematic sensitivity analysis of some key parameters. In particular, we investigate how the base-case results change as we vary the values of the probabilities of phenotypic variation $\lambda_H$ and $\lambda_L$ (Section~\ref{ResC}), the parameters $b$ and $c$ in~\eqref{initcondsdi} (Section~\ref{inits}) -- \emph{i.e.} the inverse of the initial standard deviation and the initial mean phenotype of the two populations -- and the parameters $a_H$ and $a_L$~\eqref{initcondsdi} (Section~\ref{inits2}) -- \emph{i.e.} the initial sizes of the two populations. 

We consider the nutrient concentration to be non-dimensionalised and use the dimensionless parameter values listed in Table~\ref{table1} to carry out numerical simulations of the IB model. The methods employed to numerically solve the equations of the related continuum model are described in~Appendix~\ref{nummet}.  

\subsection{\label{ResB} Base-case results}
\begin{figure}[!htbp]
\includegraphics[width=1\linewidth]{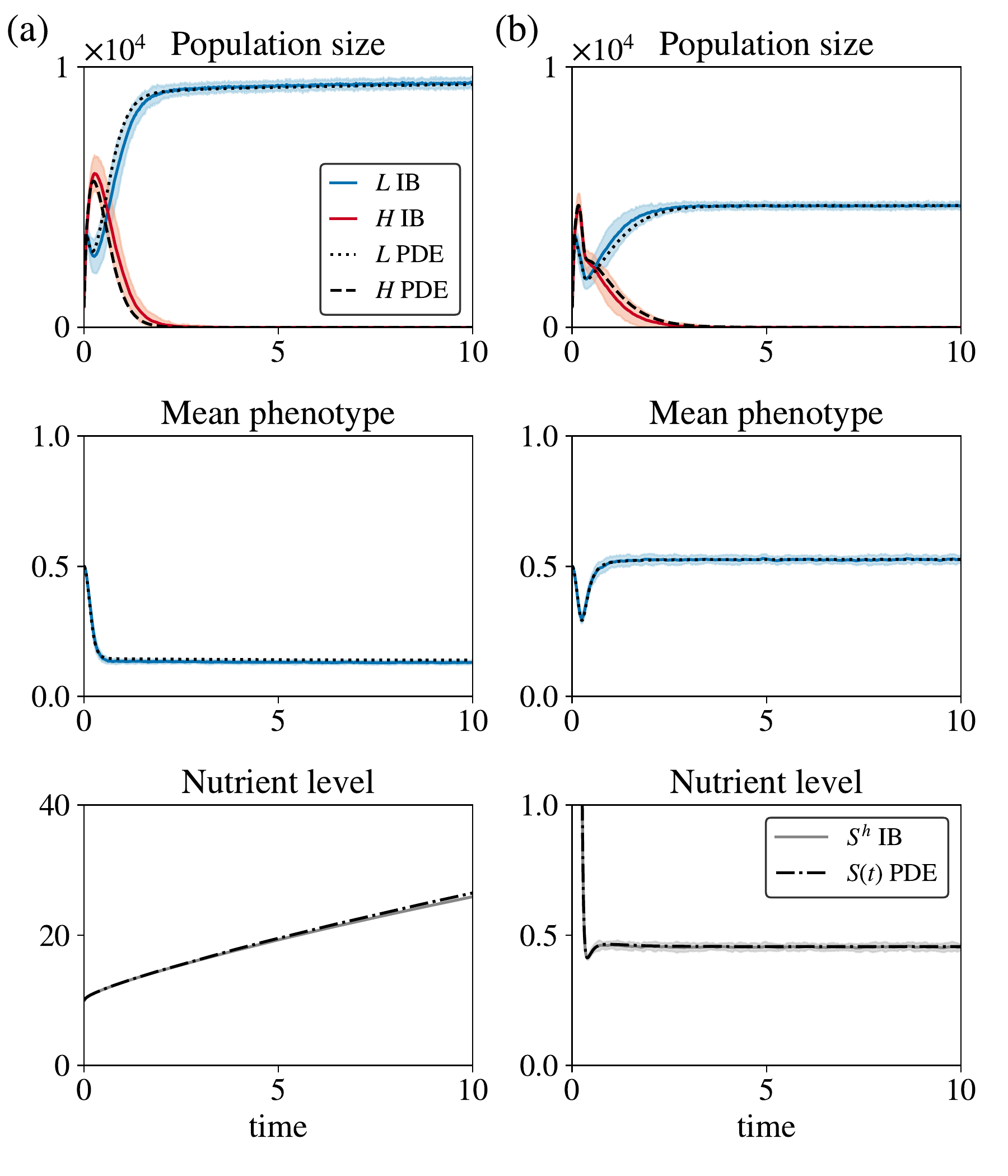}% Here is how to import EPS art
\caption{\label{fig3} {\bf Base-case results when the nutrient inflow is constant.} Comparison between numerical simulations of the IB (solid, coloured lines) and continuum (broken, black lines) models in the case where the evolution of the nutrient concentration is governed by the difference equation~\eqref{Sode} whereby the term $I^h$ is defined via~\eqref{Iconst} with $\bar{I}=10$. \textbf{(a)} Dynamics of the population sizes (top panel), mean phenotype of the surviving population (central panel) and nutrient level (bottom panel) in the case where $\theta=10^{-5}$. Here, $a_H=a_L=800$, $b = 10$ and $c=0.5$ in~\eqref{initcondsdi}, and the values of the other parameters are those listed in Table~\ref{table1} with $\lambda_H = 1$ and $\lambda_L=0.2$. The results from the IB model correspond to the average over 30 realisations and the related variance is displayed by the coloured areas surrounding the curves. \textbf{(b)} Same as \textbf{(a)} but for larger nutrient consumption, $\theta = 10^{-4}$.}
\end{figure}

%We now study the case where the nutrient concentration coevolves with the cells according to the difference equation~\eqref{Sode}. 
We first assume that the supply rate of nutrient is constant (\emph{i.e.} we define the term $I^h$ via~\eqref{Iconst}) and consider different values of the nutrient consumption rate $\theta$. The results displayed in Figure~\ref{fig3} show excellent quantitative agreement between numerical simulations of the IB and continuum models, both for relatively low and relatively high values of $\theta$. As expected, based on the results we presented in~\cite{ardavseva2019mathematical}, population $L$ outcompetes population $H$, which eventually goes extinct. Moreover, since the nutrient concentration converges to smaller equilibrium values for larger values of the nutrient consumption rate, higher values of $\theta$ correspond to decreasing equilibrium sizes of population $L$ and equilibrium values of the mean phenotype which are closer to $1$ (\emph{i.e.} the fittest phenotypic state in nutrient-scarce environments). In all cases, the phenotype distribution of the surviving population is unimodal and attains its maximum at the mean phenotype (results not shown).

\begin{figure}[!htbp]
\includegraphics[width=1\linewidth]{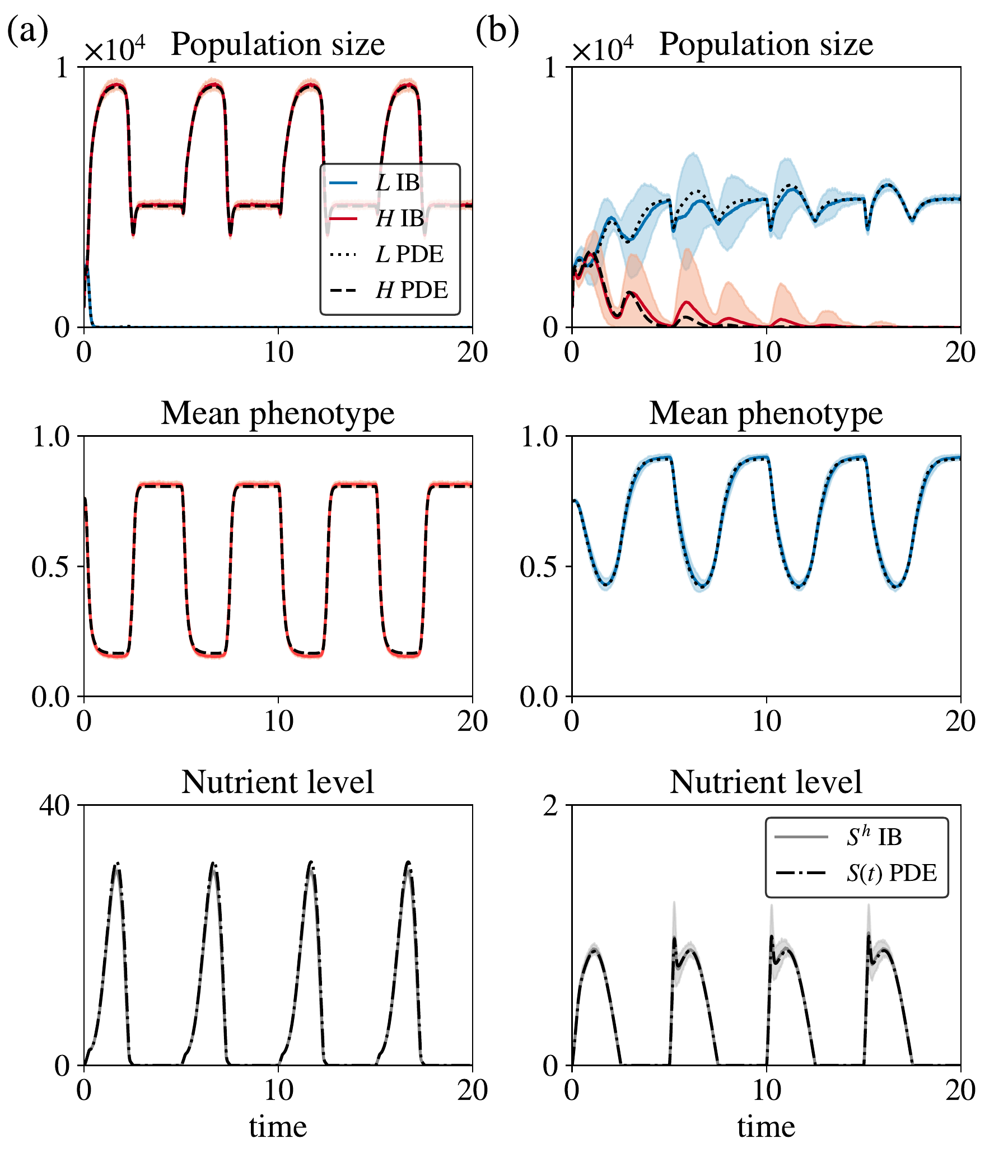}% Here is how to import EPS art
\caption{\label{fig3b} {\bf Base-case results when the nutrient inflow is periodic.} Comparison between numerical simulations of the IB (solid, coloured lines) and continuum (broken, black lines) models in the case where the evolution of the nutrient concentration is governed by the difference equation~\eqref{Sode} whereby the term $I^h$ is defined via~\eqref{Ifluc} with $A=200$ and $T=5$. \textbf{(a)} Dynamics of the population sizes (top panel), mean phenotype of the surviving population (central panel) and nutrient level (bottom panel) in the case where $\theta=2 \times 10^{-4}$. Here, $a_H=a_L=800$, $b = 10$ and $c=0.5$ in~\eqref{initcondsdi}, and the values of the other parameters are those listed in Table~\ref{table1} with $\lambda_H = 0.4$ and $\lambda_L=0.02$. The results from the IB model correspond to the average over 30 realisations and the related variance is displayed by the coloured areas surrounding the curves. \textbf{(b)} Same as \textbf{(a)} but for larger nutrient consumption, $\theta = 10^{-3}$.}
\end{figure}

We then let the supply rate of nutrient undergo periodic oscillations (\emph{i.e.} we define the term $I^h$ via~\eqref{Ifluc}) and, informed by numerical results presented in~\cite{ardavseva2019mathematical}, we consider different values of the consumption rate $\theta$ that lead to the emergence of either mild (\emph{i.e.} small-amplitude) or severe (\emph{i.e.} large-amplitude) fluctuations in the nutrient concentration $S^h$. The results displayed in Figure~\ref{fig3b} demonstrate that, both for mild and severe fluctuations in the nutrient concentration, the size and the mean phenotype of the surviving population converge to positive $T$-periodic functions. Furthermore, in agreement with the analytical results we presented in~\cite{ardavseva2019evolutionary}, the numerical results in Figure~\ref{fig3b} indicate that, when nutrient levels undergo smaller fluctuations, population $L$ survives (see Figure~\ref{fig3b}(b)). On the other hand, when nutrient levels undergo larger fluctuations, population $H$ ultimately outcompetes population $L$ (see Figure~\ref{fig3b}(a)). In both cases, the phenotype distribution of the surviving population is unimodal and attains its maximum at the mean phenotype (results not shown). Moreover, excellent agreement between numerical simulations of the IB and continuum models is observed.

The results presented in Appendix~\ref{afig} show that analogous conclusions hold in the simplified scenario where the concentration of nutrient is prescribed and does not coevolve with the cells. 

\subsection{\label{ResC} Sensitivity analysis of the probabilities of phenotypic variation}
\begin{figure}[!htbp]
\includegraphics[width=1\linewidth]{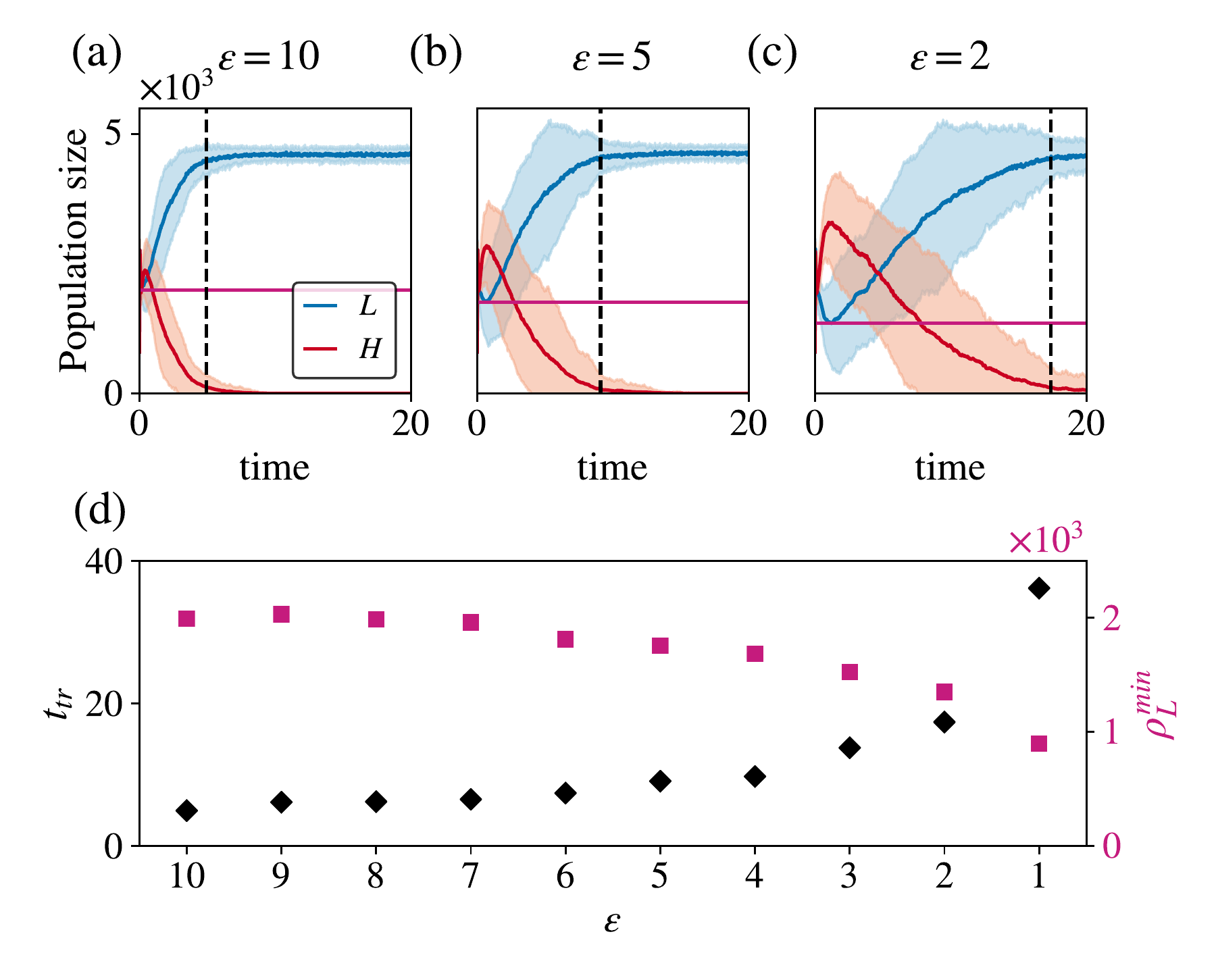}
\caption{\label{fig4} {\bf Emergence of longer transient intervals for lower probabilities of phenotypic variation.} {\bf (a)-(c)}. Numerical simulations of the IB model in the case where the probabilities of phenotypic variation $\lambda_H$ and $\lambda_L$ are defined via~\eqref{deflambdai} with $\Lambda_H=0.05$, $\Lambda_L=0.02$, and $\varepsilon = 10$ (panel {\bf (a)}) or $\varepsilon = 5$ (panel {\bf (b)}) or $\varepsilon = 2$ (panel {\bf (c)}). The black dashed lines highlight the time $t_{tr}$ such that $\rho_L^{t_{f}} - \rho_L^{t_{tr}} < 100$, while the solid pink lines highlight the value of $\rho_L^{min} := \displaystyle{\min_h \rho_L^h}$. These results correspond to the average over 30 realisations and the related variance is displayed by the coloured areas surrounding the curves. {\bf (d)}. Plots of $t_{tr}$ (black diamonds) and $\rho_L^{min}$ (pink squares) as functions of $\varepsilon \in \{1, \ldots, 10\}$. The evolution of the nutrient concentration is governed by the difference equation~\eqref{Sode}, whereby the term $I^h$ is defined via~\eqref{Iconst} with $\bar{I}=10$. Here, $a_H=a_L=800$, $b = 1000$ and $c=0.5$ in~\eqref{initcondsdi}, and the values of the other parameters are those listed in Table~\ref{table1} with $\theta= 10^{-3}$.}
\end{figure}

Based on the analytical results presented in~\cite{ardavseva2019evolutionary} for a simplified continuum model, we expect smaller values of $\lambda_H$ and $\lambda_L$ (\emph{i.e.} the probabilities of phenotypic variation) to correlate with longer transient intervals in the dynamics of the sizes of the two cell populations. To test this hypothesis, we focus on the case where the supply rate of nutrient is constant (\emph{i.e.} when the term $I^h$ is defined via~\eqref{Iconst}). We carry out numerical simulations of the IB model assuming
\begin{equation}
\label{deflambdai}
\lambda_i = \varepsilon \ \Lambda_i,
\end{equation}
with $\Lambda_i$ fixed and $\varepsilon \in \{1, \ldots, 10\}$. As summarised by the plots in Figure~\ref{fig4}, smaller values of $\varepsilon$ bring about longer transient intervals (\emph{i.e.} larger values of $t_{tr}$ in Figure~\ref{fig4}(d)) during which the two populations coexist before population $L$ ultimately out-competes population $H$. 

The results displayed in Figure~\ref{fig4}(a)--(c) indicate that the size of population $L$ decreases during the transient. Moreover, longer transients correlate with lower minimum values of the size of population $L$ (\emph{i.e.} smaller $\rho_L^{min}$ in Figure~\ref{fig4}(d)), which makes the possibility of stochastic effects, associated with small population sizes, more likely to come into play. This suggests that lower probabilities of phenotypic variation may create conditions for the emergence of differences between predictions of the IB and continuum models. 

\begin{figure}[h!]
\includegraphics[width=1\linewidth]{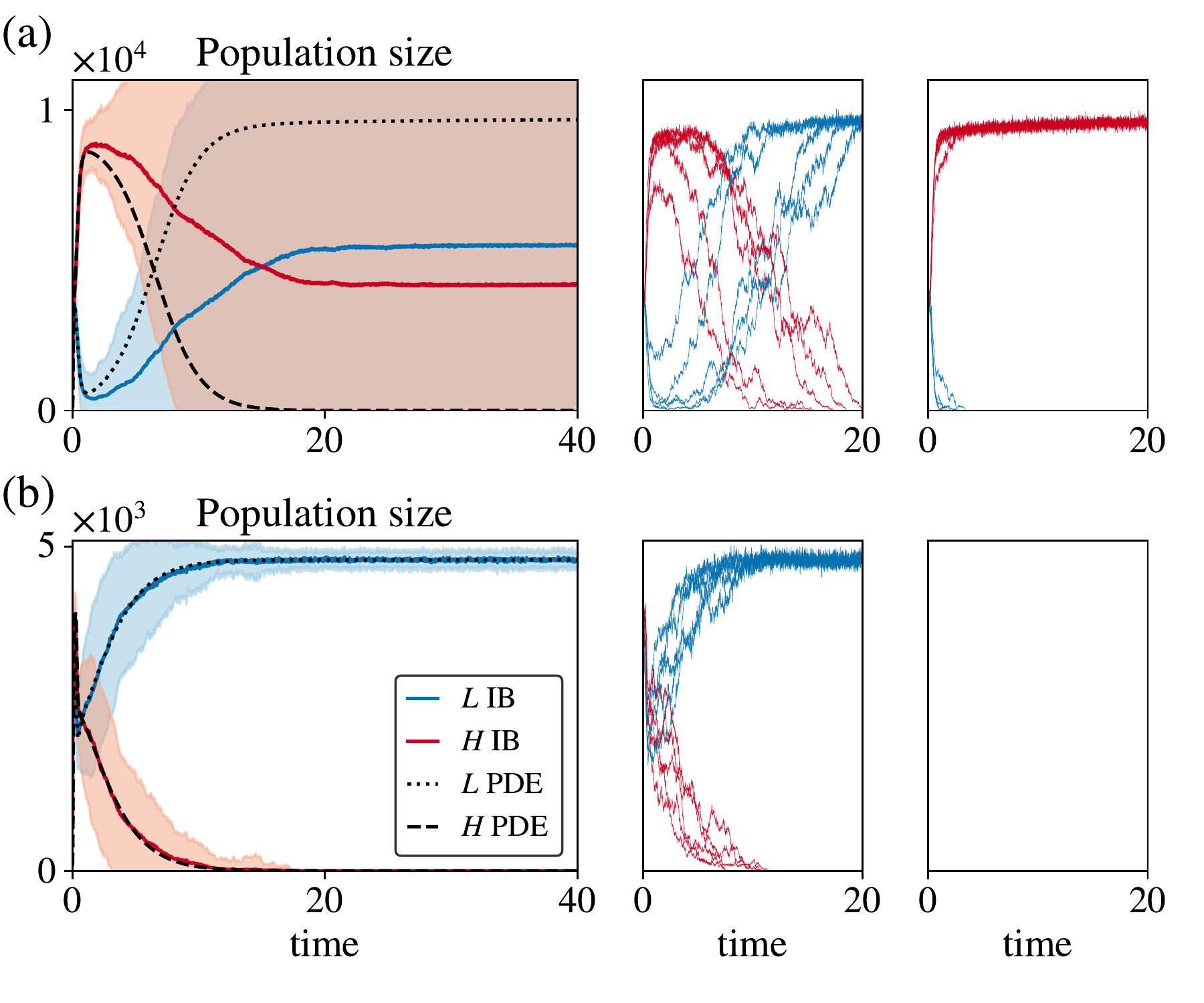}% Here is how to import EPS art
\caption{\label{Fig5}  {\bf Sensitivity analysis of the probabilities of phenotypic variation.} \textbf{(a)} Comparison between numerical simulations of the IB  (solid, coloured lines in the left panel) and continuum (broken, black lines in the left panel) models under the parameter setting of Figure~\ref{fig3}(a) but with $\lambda_H = 0.05$ and $\lambda_L=0.02$. The results from the IB model correspond to the average over 30 realisations and the related variance is displayed by the coloured areas surrounding the curves. The plots in the central and right panels show the dynamics of the sizes of the two populations for single realisations of the IB model that match with (central panel) or differ from (right panel) numerical solutions of the continuum model. \textbf{(b)} Same as \textbf{(a)} for the parameter setting of Figure~\ref{fig3}(b) but with $\lambda_H = 0.05$ and $\lambda_L=0.02$.}
\end{figure}

To investigate this further, we compare numerical simulations of the IB model with numerical solutions of the continuum model in the setting of Figure~\ref{fig3} (\emph{i.e.} defining the term $I^h$ via~\eqref{Iconst} and considering different values of $\theta$) but using lower values of the parameters $\lambda_H$ and $\lambda_L$. The results, summarised in Figure~\ref{Fig5}, demonstrate that while excellent quantitative agreement between numerical simulations of the IB model and numerical solutions of the continuum model is obtained for relatively large values of $\theta$ (see Figure~\ref{Fig5}(b)), significant differences in the behaviour of the two models can be observed for relatively low values of $\theta$ (see Figure~\ref{Fig5}(a)). 

This is because, when lower values of $\lambda_H$ and $\lambda_L$ are considered, relatively small $\theta$ correspond to a longer initial phase of cell dynamics during which the size of population $L$ decays and the size of population $H$ grows. After this initial phase, the numerical solutions of the continuum model exhibit trend inversion, with the size of population $L$ converging to a stable positive value and the size of population $H$ decaying to zero. On the other hand, numerical simulations of the IB model demonstrate that, {\it ceteris paribus}: for some realisations population $L$ can recover from the initial decay (see central panel of Figure~\ref{Fig5}(a)) -- in these cases an excellent quantitative match between the outcomes of the two models is observed; there are realisations whereby, due to stochastic effects, the aftermath of the initial phase of cell dynamics is the extinction of population $L$ and the survival of population $H$ (see right panel of Figure~\ref{Fig5}(a)). As a result, on average, the IB model predicts coexistence between the two cell populations, whereas the continuum model predicts extinction of population $H$.

Differences between the discrete and the continuum model are also observed when the supply rate of nutrient undergoes periodic oscillations (\emph{i.e.} when the term $I^h$ is defined via~\eqref{Ifluc}) and different values of $\theta$ are considered, provided that lower values of $\lambda_H$ and $\lambda_L$ are chosen (results not shown). In this case, for values of $\theta$ leading to the emergence of severe fluctuations in the nutrient level (\emph{i.e.} when population $H$ is ultimately selected according to the continuum model), there is an excellent quantitative agreement between the two models. On the other hand, for values of $\theta$ leading to the emergence of mild fluctuations in nutrient levels (\emph{i.e.} when the continuum model predicts that population $L$ will ultimately be selected after an initial phase of population size contraction), there are realisations of the IB model in which population $L$ is outcompeted by population $H$ and, on average, coexistence between the two cell populations occurs.

\subsection{\label{inits} Sensitivity analysis of the initial standard deviation and the initial mean phenotype}
\begin{figure*}[!htb]
\includegraphics[width=1\linewidth]{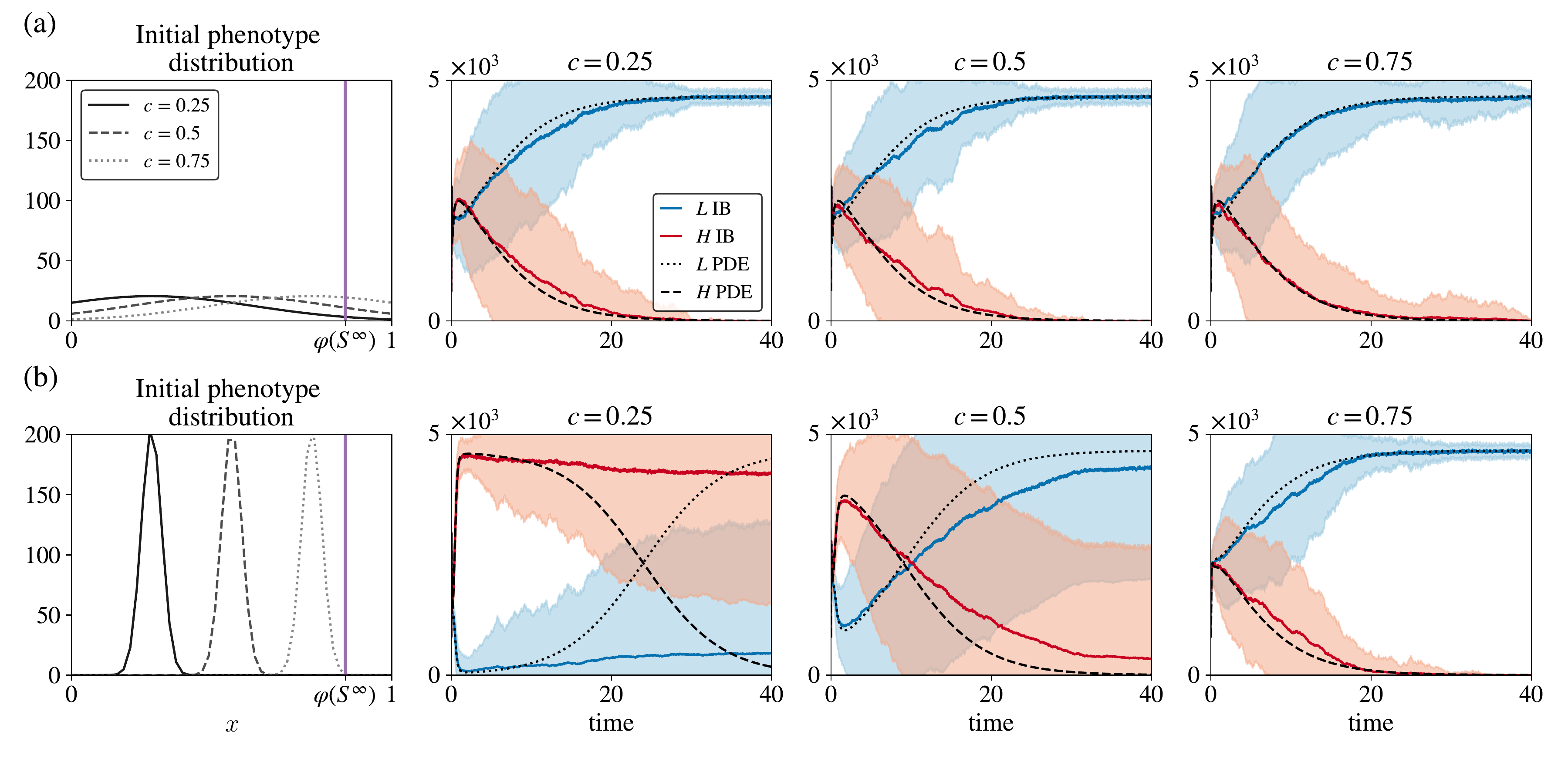}% Here is how to import EPS art
\caption{\label{fig6} {\bf Sensitivity analysis of the initial mean phenotype and the initial standard deviation when the nutrient inflow is constant.} \textbf{(a)} Comparison between numerical simulations of the IB (solid, coloured lines in panels 2--4) and continuum (broken, black lines in panels 2--4) models in the case where the initial phenotype distributions of the two populations are defined as shown by the plots in the first panel, corresponding to different values of $c$ in~\eqref{initcondsdi}. The purple line in the first panel highlights the equilibrium value of the fittest phenotypic state $\varphi{(S^{\infty})}$, which is computed by substituting into~\eqref{def:phi} the long-time limit $S^{\infty}$. The evolution of $S^h$ is governed by the difference equation~\eqref{Sode}, whereby the term $I^h$ is defined via~\eqref{Iconst} with $\bar{I}=10$. Here, $a_H=a_L=800$ and $b = 10$ in~\eqref{initcondsdi}, and the values of the other parameters are those listed in Table~\ref{table1} with $\lambda_H = 0.05$, $\lambda_L=0.02$ and $\theta = 10^{-3}$. The results from the IB model correspond to the average over 30 realisations and the related variance is displayed by the coloured areas surrounding the curves. \textbf{(b)} Same as \textbf{(a)} but for $b = 1000$.}
\end{figure*}

Based on analytical results presented in~\cite{ardavseva2019evolutionary} for a simplified continuum model, in the case where the nutrient concentration coevolves with the cells according to the difference equation~\eqref{Sode} and the supply rate $I^h$ is defined via~\eqref{Iconst}, we anticipate longer transient intervals in the dynamics of the sizes of the two cell populations in the presence of both small initial standard deviations, $\sigma^0_{H,L}$, and large distances between the initial mean phenotypes, $\mu^0_{H,L}$, and the equilibrium value of the fittest phenotypic state $\varphi{(S^{\infty})}$, which is computed by substituting the long-time limit $S^{\infty}$ of the nutrient concentration into~\eqref{def:phi}. Since the results presented in Section~\ref{ResC} demonstrate that longer transient intervals may enhance the stochastic effects associated with small population sizes, we expect that larger values of $|\mu^0_H - \varphi(S^{\infty})|$ and $|\mu^0_L - \varphi(S^{\infty})|$, along with smaller values of $\sigma^0_H$ and $\sigma^0_L$, will increase the likelihood of observing differences between numerical simulations of the IB and continuum models. 

To test this hypothesis, we first suppose the nutrient supply rate to be constant and we carry out numerical simulations for different values of the parameters $b$ and $c$ in~\eqref{initcondsdi}. We recall that larger values of $b$ correlate with lower $\sigma^0_H$ and $\sigma^0_L$, and in the setting considered here, lower values of $c$ correspond to higher $|\mu^0_H - \varphi(S^{\infty})|$ and $|\mu^0_L - \varphi(S^{\infty})|$ (\emph{i.e.} less fit initial mean phenotypes). The plots presented in Figure~\ref{fig6}(a) reveal excellent quantitative agreement between numerical simulations of the IB and continuum models for sufficiently large values of $\sigma^0_H$ and $\sigma^0_L$, regardless of the values of $|\mu^0_H - \varphi(S^{\infty})|$ and $|\mu^0_L - \varphi(S^{\infty})|$ (\emph{i.e.} independently of the value of $c$). On the other hand, and consistent with our expectations, the numerical results presented in Figure~\ref{fig6}(b) show that, for sufficiently small values of $\sigma^0_H$ and $\sigma^0_L$, higher $|\mu^0_H - \varphi(S^{\infty})|$ and $|\mu^0_L - \varphi(S^{\infty})|$ (\emph{i.e.} lower values of $c$) correlate with longer transients during which stochastic effects can lead to the emergence of differences between the cell dynamics produced by the two models.

\begin{figure*}[htb!]
\includegraphics[width=1\linewidth]{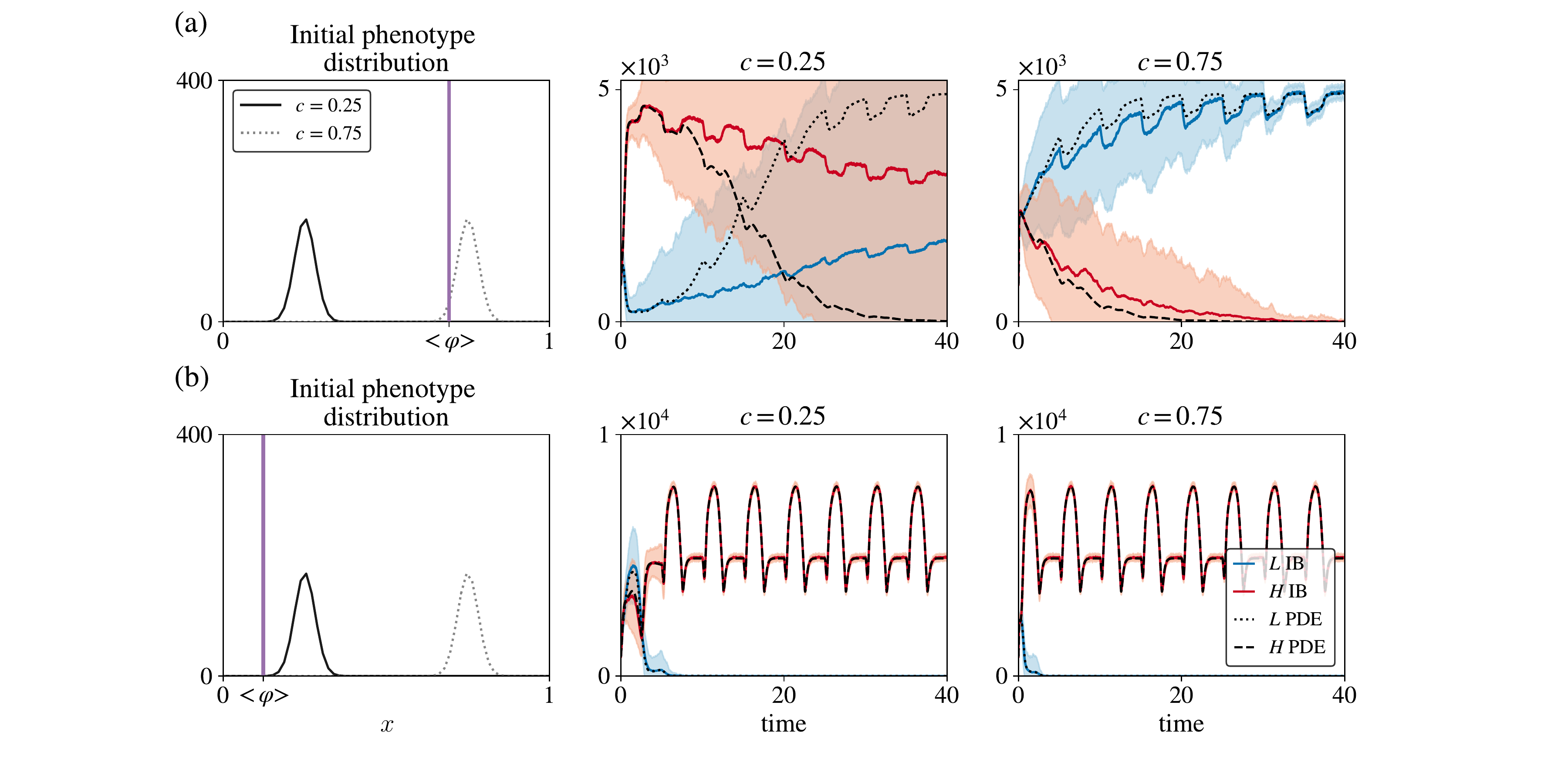}% Here is how to import EPS art
\caption{\label{fig8} {\bf Sensitivity analysis of the initial mean phenotype when the nutrient inflow is periodic.} \textbf{(a)} Comparison between numerical simulations of the IB (solid, coloured lines in the central and right panel) and continuum (broken, black lines in the central and right panel) models in the case where the initial phenotype distributions of the two populations are the same and both defined as shown by the plots in the left panel, which correspond to different values of the parameter $c$ in~\eqref{initcondsdi}. The purple line in the first panel highlights the value of the quantity $<\varphi> $ defined according to~\eqref{phiunc}. The evolution of $S^h$ is governed by the difference equation~\eqref{Sode}, whereby the term $I^h$ is defined via~\eqref{Ifluc} with $A=30$ and $T=5$. Numerical simulations are carried out assuming $a_H=a_L=800$ and $b = 1000$ in~\eqref{initcondsdi}, and using the parameter values listed in Table~\ref{table1} with $\lambda_H = 0.05$, $\lambda_L=0.02$ and $\theta = 10^{-3}$. The results from the IB model correspond to the average over 30 realisations and the related variance is displayed by the coloured areas surrounding the curves. \textbf{(b)} Same as \textbf{(a)} but for $\theta = 5\times10^{-5}$.} 
\end{figure*}

We now suppose that the nutrient supply rate undergoes periodic oscillations and perform numerical simulations for different values of the parameter $c$, which correspond to different values of the quantities $|\mu^0_H - <\varphi>|$ and $|\mu^0_L - <\varphi>|$, where 
\begin{equation}
\label{phiunc}
<\varphi> := \frac{1}{2} \left(\min_{t_h \in [0,T]} \tilde{S}^h + \max_{t_h \in [0,T]} \tilde{S}^h\right)
\end{equation}
with $\tilde{S}^h$ being the positive $T$-periodic function to which $S^h$ converges as $h \to \infty$. In the setting considered here, smaller values of $c$ correspond to higher $|\mu^0_H - <\varphi>|$ and $|\mu^0_L - <\varphi>|$ (\emph{i.e.} less fit initial mean phenotypes). The results presented in Figure~\ref{fig8}(b) indicate that excellent quantitative agreement is observed between numerical simulations of the IB and continuum models when the consumption rate $\theta$ is such that the nutrient level undergoes severe fluctuations (\emph{i.e.} when population $H$ is ultimately selected according to the continuum model), regardless of the values of $|\mu^0_H - <\varphi>|$ and $|\mu^0_L - <\varphi>|$ (\emph{i.e.} independently of the value of $c$). On the other hand, the results presented in Figure~\ref{fig8}(a) show that, when $\theta$ is such that the nutrient level undergoes mild fluctuations (\emph{i.e.} when the continuum model predicts population $L$ to be ultimately selected after an initial phase of population size contraction), good quantitative agreement between numerical simulations of the IB and continuum models is observed only if $|\mu^0_L - <\varphi>|$ and $|\mu^0_H - <\varphi>|$ are sufficiently small (\emph{i.e.} only if $c$ is sufficiently large). Indeed, larger values of these distances correlate with longer transients during which stochastic effects may drive discrepancies between the cell dynamics of the two models.

\subsection{Sensitivity analysis of the initial population sizes \label{inits2}}
Motivated by the numerical results presented in Section~\ref{inits}, we hypothesise that differences between numerical simulations of the IB and continuum models, which are observed for sufficiently large values of $|\mu^0_i - \varphi(S^\infty)|$ (\emph{i.e.} sufficiently small $c$) and sufficiently small values of $\sigma^0_i$ (\emph{i.e.} sufficiently high $b$), will be amplified when smaller initial sizes of population $L$ are considered and the initial total number of cells is held fixed. Indeed, lower values of $\rho_L^0$ may exaggerate stochastic effects associated with small population sizes during the initial phase of the cell dynamics (\emph{i.e.} when the size of population $L$ decays). To test this hypothesis, we focus on the case where the nutrient inflow rate is constant and carry out numerical simulations for which the parameters $a_H$ and $a_L$ in~\eqref{initcondsdi} are related as follows
\begin{equation}
\label{defahl}
a_H = \nu Z \quad \text{and} \quad a_L = (1-\nu) Z,
\end{equation}
with $Z$ fixed and for increasing values of $0 < \nu < 1$.

\begin{figure}[h!]
\includegraphics[width=1\linewidth]{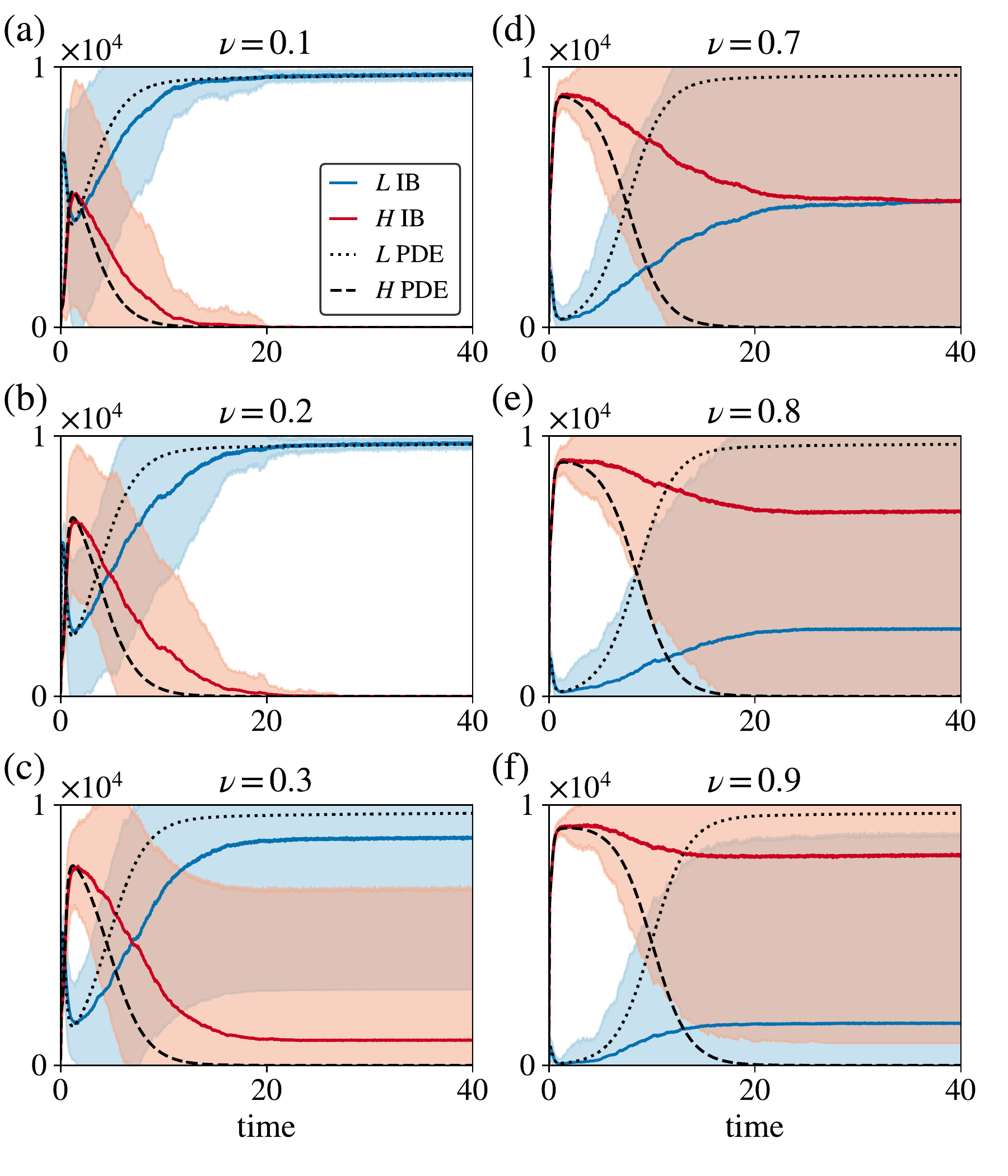}% Here is how to import EPS art
\caption{\label{Fig7} {\bf Sensitivity analysis of the initial population sizes.}  \textbf{(a)} Comparison between numerical simulations of the IB (solid, coloured lines) and continuum (broken, black lines) models in the case where $a_i$ in~\eqref{initcondsdi} is defined via~\eqref{defahl} with $Z=800$ and $\nu=0.1$. The evolution of the nutrient concentration is governed by the difference equation~\eqref{Sode}, whereby the term $I^h$ is defined via~\eqref{Iconst} with $\bar{I}=10$. Numerical simulations are carried out assuming $b = 1000$ and $c = 0.5$ in~\eqref{initcondsdi}, and using the parameter values listed in Table~\ref{table1} with $\lambda_H = 0.05$, $\lambda_L=0.02$ and $\theta = 10^{-3}$. The results from the IB model correspond to the average over 30 realisations and the related variance is displayed by the coloured areas surrounding the curves. \textbf{(b)--(f)} Same as \textbf{(a)} but for $\nu=0.2$ (panel \textbf{(b)}), $\nu=0.3$ (panel \textbf{(c)}), $\nu=0.7$ (panel \textbf{(d)}), $\nu=0.8$ (panel \textbf{(e)}), $\nu=0.9$ (panel \textbf{(f)}).}
\end{figure}

The results presented in Figure~\ref{Fig7} show that, higher values of $\nu$ lead to longer transient intervals, during which the two populations coexist. For all admissible values of $\nu$, the solutions of the continuum model are such that the size of population $L$ evolves to a stable positive value and population $H$ becomes extinct. By contrast, numerical simulations of the IB model reveal that for $\nu$ sufficiently large, on average, stable coexistence between the two cell populations occurs at long times. Moreover, the size of population $H$ may undergo small stochastic fluctuations about a stable positive value that is larger than that about which the size of population $L$ fluctuates -- \emph{i.e.} the mean size of population $H$ is higher than the mean size of population $L$. 

Analogous results pertain when a periodic nutrient inflow defined via~\eqref{Ifluc} is considered, provided that values of $\theta$ leading to the emergence of mild fluctuations in the nutrient level are chosen (\emph{i.e.} when the continuum model predicts population $L$ to be ultimately selected after an initial phase of population size contraction) along with sufficiently high $|\mu^0_L - <\varphi>|$ and $|\mu^0_H - <\varphi>|$ (results not shown).

\section{\label{conclusion} Conclusions}
We have developed a stochastic IB model for the evolutionary dynamics of two competing phenotype-structured cell populations that are exposed to time-varying nutrient levels and undergo spontaneous, heritable phenotypic variations with different probabilities. We have formally derived the deterministic continuum counterpart of this model and carried out a systematic comparison between numerical simulations of the IB and continuum models. 

We have presented base-case results that demonstrate an excellent quantitative match between the outcomes of the two models. These results agree with our previously published analytical and numerical results for related deterministic continuum models~\citep{ardavseva2019evolutionary, ardavseva2019mathematical}. Moreover, we investigated the importance of stochastic effects in driving differences between the predictions made by the two models and how these cannot be captured by the deterministic continuum model. The results indicate that stochastic effects associated with small population sizes, which are crucial in population bottlenecks, can lead to significant differences between the two models. In particular, these differences arise in the presence of low probabilities of phenotypic variation, and are more apparent when the two populations are characterised by less fit initial mean phenotypes and smaller initial levels of phenotypic heterogeneity. When there is agreement between the two modelling approaches, this is also dependent on the initial proportions of the two populations.

The generality of our assumptions make the discrete modelling framework considered here applicable to a broad range of asexual organisms exposed to dynamically changing environments. Such a modelling framework, along with the related method to formally derive corresponding continuum models, can be easily extended to incorporate the effects of additional biological aspects related to spatial structure, such as cell movement, inter-cellular spatial interactions, nutrient diffusion and the presence of multiple sources of nutrient distributed across the spatial domain. These extensions will enable a more biologically relevant exploration of the scenarios under which stochastic effects may result in discrepancies between the predictions made by discrete stochastic models and those made by their deterministic continuum limits. This will ultimately help disentangle the impact of, different sources of, stochasticity on the emergence of spatio-temporal evolutionary patterns in a variety of living systems~\citep{robertson2015impact, thuiller2007stochastic}.  

% The \nocite command causes all entries in a bibliography to be printed out
% whether or not they are actually referenced in the text. This is appropriate
% for the sample file to show the different styles of references, but authors
% most likely will not want to use it.
%\nocite{*}
\begin{acknowledgments}
AA is supported by funding from the Engineering and Physical Sciences Research Council (EPSRC) and the Medical Research Council (MRC) (grant no. EP/L016044/1) and in part by the Moffitt Cancer Center PSOC, NIH/NCI (grant no. U54CA193489). RG and ARAA are supported by Physical Sciences Oncology Network (PSON) grant from the National Cancer Institute (grant no. U54CA193489) as well as the Cancer Systems Biology Consortium grant from the National Cancer Institute (grant no. U01CA23238). ARAA and RG would also like to acknowledge support from the Moffitt Cancer Center of Excellence for Evolutionary Therapy.
\end{acknowledgments}

\appendix
\section{\label{formalderiv} Formal derivation of the continuum model given by~\eqref{eS1}}
Using a method analogous to that employed in~\cite{chisholm2016evolutionary, stace2019discrete}, we show that the system of non-local PDEs~\eqref{eS1} can be formally derived as the appropriate continuum limit of our discrete model. 

In the case where the dynamics of the cells is governed by the rules described in Section~\ref{model}, the principle of mass balance gives the following difference equations
\begin{multline} 
    n_{i,j}^{h+1} = \left\{2 \, \tau \, p(x_j,S^h) +  \left[1 - \tau \left(p(x_j,S^h) + d\rho^h\right) \right]  \right\} \nonumber \\ \times  \Big[\frac{\lambda_i}{2} n_{i,j+1}^h + \frac{\lambda_i}{2} n_{i,j-1}^h + \left(1-\lambda_i\right) n_{i,j}^h \Big], \nonumber
\end{multline}
for $i\in\left\{H,L\right\}$, which can be rewritten as
\begin{multline}\label{eq:intera1}
    n_{i,j}^{h+1} = (1 + \tau \ p(x_j,S^h) - \tau \ d \ \rho^h) \Big[\frac{\lambda_i}{2} n_{i,j+1}^h + \\ \frac{\lambda_i}{2} n_{i,j-1}^h + (1-\lambda_i) n_{i,j}^h \Big].
\end{multline}
Using the fact that the following relations hold for $\tau$ and $\chi$ sufficiently small
\begin{align*}
    & t_h \approx t, \quad t_{h+1} \approx t + \tau,  \quad x_j \approx x, \quad x_{j \pm 1} \approx x \pm \chi, \\
    &n_{i,j}^h \approx n_i(x,t), \quad S^h \approx S(t), \\
    &n_{i,j}^{h+1} \approx n_i(x, t+\tau), \quad n_{i,j\pm1}^h \approx n_i(x\pm\chi, t),\\
     &\rho^h_i \approx \rho_i(t) := \int_{0}^{1} n_i(x,t) \; \textup{d}x,\\
    &\rho^h \approx \rho(t) := \int_{0}^{1} n_H(x,t) \; \textup{d}x +\int_{0}^{1} n_L(x,t) \; \textup{d}x,
\end{align*}
equation~ \eqref{eq:intera1} can be formally rewritten in the approximate form
\begin{multline}\label{e:deriv1}
    n_i(x,t+\tau) = \Big(1+\tau R(x, S(t),\rho(t))\Big) \Big[\frac{\lambda_i}{2} n_i(x+\chi,t) + \\ \frac{\lambda_i}{2} n_i(x-\chi,t) + (1-\lambda_i) n_i(x,t) \Big],
\end{multline}
with $R(x,S(t),\rho(t)) := p(x,S(t)) - d\rho(t)$. If the function $n_i(x,t)$ is twice continuously differentiable with respect to the variable $x$, for $\chi$ sufficiently small we can use the Taylor expansions
%Assuming that $n_i \in C^2 (\mathbb{R} \times \mathbb{R}_{>=0})$, we can approximate the terms $n_i(x,t+\tau)$, $n_i(x + \chi,t)$ and $n_i(x-\chi,t)$ by their second order Taylor expansions about the point $(x,t)$:
\begin{equation}\label{taylorexp}
n_i(x\pm \chi,t) = n_i \pm \chi \frac{\partial n_i}{\partial x} + \frac{\chi^2}{2} \frac{\partial^2 n_i}{\partial x^2} + h.o.t. \ ,
\end{equation}
where $n_i \equiv n_i(x,t)$. Substituting~\eqref{taylorexp} into~\eqref{e:deriv1} and dividing both sides of the resulting equation by $\tau$, after a little algebra we find
\begin{multline*}
%\label{e:deriv2}
    \frac{n_i(x,t+\tau) - n_i(x,t)}{\tau} = R(x, S(t),\rho(t)) n_i(x,t) \\ + \frac{\lambda_i \chi^2}{2 \tau} \frac{\partial^2 n_i(x,t)}{\partial x^2} \\ 
    + R(x, S(t),\rho(t)) \frac{\lambda_i \chi^2}{2} \frac{\partial^2 n_i(x,t)}{\partial x^2} + h.o.t. \ .
\end{multline*}
If, in addition, the function $n_i(x,t)$ is continuously differentiable with respect to the variable $t$, letting $\tau \to 0$ and $\chi \to 0$ in such a way that condition~\eqref{asymatdr} is met, from the latter equation we formally obtain
$$
\frac{\partial n_i(x,t)}{\partial t} = \beta_i \frac{\partial^2 n_i(x,t)}{\partial x^2} \; + \; R\big(x,S(t),\rho(t)\big) \, n_i(x,t),
$$
which gives the system of non-local PDEs~\eqref{eS1}. Finally, the zero-flux boundary conditions~\eqref{BCs} follow from the fact that the attempted phenotypic variation of a cell is aborted if it requires moving into a phenotypic state that does not belong to the interval $[0,1]$.

\section{Details of numerical simulations of the continuum model}
\label{nummet}
To construct numerical solutions of the system of non-local PDEs~\eqref{eS1} posed on $(0,1) \times (0,t_f]$, and subject both to the zero-flux boundary conditions~\eqref{BCs} and to the continuum analogue of the initial condition~\eqref{initcondsdi}, \emph{i.e.}
\begin{equation} \label{initconds}
n_{i}(x,0) = a_i \left(\frac{b}{2\pi} \right)^{\frac{1}{2}} \exp \left[-\frac{b}{2} (x-c)^2 \right], \quad 
\end{equation}
with $i \in \{H, L\}$, we use a uniform discretisation of the interval $(0,1)$ as the computational domain of the independent variable $x$, and we discretise the time interval $(0,t_f]$ with the uniform step $\Delta t =0.0001$. The method for constructing numerical solutions is based on a three-point finite difference explicit scheme for the diffusion terms and an explicit finite difference scheme for the reaction term~\citep{leveque2007finite}. Moreover, the differential equation~\eqref{SodeCont}, which is subject to the initial condition $S(0)=10$ and complemented with the continuum analogues of the alternative definitions of the term $I^h$ that are specified in the main body of the paper, is solved numerically by using an explicit Euler method with step $\Delta t$. Given the values of the parameter $\tau$, $\chi$, $\lambda_H$ and $\lambda_L$ of the IB model, the values of the parameters $\beta_H$ and $\beta_L$ are defined so that condition~\eqref{asymatdr} is met. The other parameter values are chosen to be coherent with those used to carry out numerical simulations of the IB model, which are specified in the main body of the paper.
%and ODE \eqref{SodeCont}

\section{Base-case results in the case where the nutrient concentration is prescribed}\label{afig}
We carry out preliminary numerical simulations in the case where, instead of being the solution of the difference equation~\eqref{Sode}, the nutrient concentration is prescribed and given by
\begin{equation}\label{Sgiven}
S^h := M + A \sin \left(\frac{2 \pi t_h}{T} \right),
\end{equation}
where $M > 0$ is the mean nutrient level, and the parameter $0 \leq A \leq M$ models the semi-amplitude of possible oscillations of the nutrient level, which have period $T>0$. We fix the values of $M$ and $T$ and consider three different values of $A$ that correspond to distinct environmental regimes: constant nutrient level (\emph{i.e.} no oscillations), mild nutrient fluctuations (\emph{i.e.} small-amplitude oscillations) and severe nutrient fluctuations (\emph{i.e.} large-amplitude oscillations).

\begin{figure}[h!]
\includegraphics[width=1\linewidth]{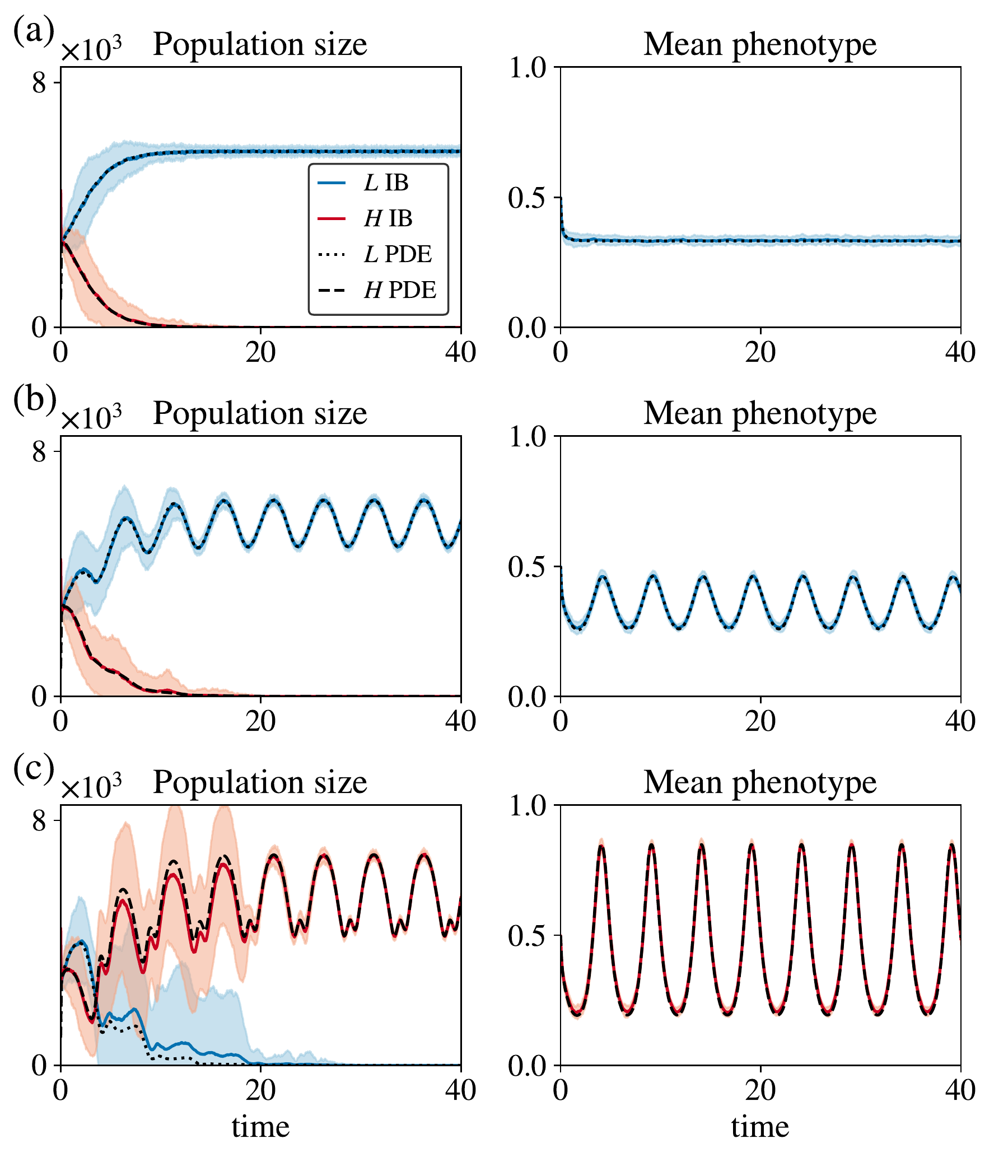}
\caption{\label{fig2} {\bf Base-case results when the nutrient concentration is prescribed.} Comparison between numerical simulations of the IB (solid, coloured lines) and continuum (broken, black lines) models in the case where the nutrient concentration is prescribed and defined via~\eqref{Sgiven}. \textbf{(a)} Dynamics of the population sizes (left column) and the mean phenotype of the surviving population (right column) in the case where $M=1$, $T=5$ and $A=0$ in~\eqref{Sgiven}. Here, $a_H=a_L=800$, $b = 10$ and $c=0.5$ in~\eqref{initcondsdi}, and the values of the other parameters are those listed in Table~\ref{table1} with $\lambda_H = 0.05$ and $\lambda_L=0.02$. The results from the IB model correspond to the average over 30 realisations and the related variance is displayed by the coloured areas surrounding the curves. \textbf{(b)--(c)} Same as \textbf{(a)} but for $A=0.5$ (row \textbf{(b)}) and $A=1$ (row \textbf{(c)}).}
\end{figure}

The results presented in Figure~\ref{fig2} show that, for all values of $A$ considered, there is an excellent quantitative match between the  numerical simulations of the IB and continuum models. In agreement with the analytical results that we presented in~\cite{ardavseva2019evolutionary}, when the nutrient concentration is constant, population $L$ outcompetes population $H$ (see Figure~\ref{fig2}(a)). The same outcome is observed in the presence of mild nutrient fluctuations (see Figure~\ref{fig2}(b)). By contrast, population $L$ is outcompeted by population $H$ when severe nutrient fluctuations occur (see Figure~\ref{fig2}(c)). In all cases, the phenotype distribution of the surviving population is unimodal and attains its maximum at the mean phenotype (results not shown). Moreover, when the nutrient level is constant, the size and the mean phenotype of the surviving population converge to stable values. On the other hand, in the presence of $T$-periodic nutrient fluctuations, the size and mean phenotype of the surviving population converge to $T$-periodic functions.

\bibliography{apssamp}% Produces the bibliography via BibTeX.

\end{document}